# Observation of topological edge states in thermal diffusion

Hao Hu[1], Song Han[1], Yihao Yang[2,3,4], Dongjue Liu[1], Haoran Xue[5], Gui-Geng Liu[5], Zheyu Cheng[5], Qi Jie Wang[1,5,8], Shuang Zhang[6,7], Baile Zhang[5,8], and Yu Luo[1]

[1] School of Electrical and Electronic Engineering, Nanyang Technological University, 50 Nanyang Avenue, Singapore 639798, Singapore.
[2] Interdisciplinary Center for Quantum Information, State Key Laboratory of Modern Optical Instrumentation, College of Information Science and Electronic Engineering, Zhejiang University, Hangzhou 310027, China.
[3] ZJU-Hangzhou Global Science and Technology Innovation Center, Key Lab. of Advanced Micro/Nano Electronic Devices & Smart Systems of Zhejiang, Zhejiang University, Hangzhou 310027, China.
[4] International Joint Innovation Center, ZJU-UIUC Institute, Zhejiang University, Haining 314400, China.
[5] Division of Physics and Applied Physics, School of Physical and Mathematical Sciences, Nanyang Technological University, 21 Nanyang Link, Singapore 637371, Singapore.
[6] Department of Physics, University of Hong Kong, Hong Kong, China.
[7] Department of Electrical & Electronic Engineering, University of Hong Kong, China.
[8] Centre for Disruptive Photonic Technologies, Nanyang Technological University, Singapore 637371, Singapore.

**The topological band theory predicts that bulk materials with nontrivial topological phases support topological edge states. This phenomenon is universal for various wave systems and has been widely observed for electromagnetic and acoustic waves. Here, we extend the notion of band topology from wave to diffusion dynamics. Unlike the wave systems that are usually Hermitian, the diffusion systems are anti-Hermitian with purely imaginary eigenvalues corresponding to decay rates. Via direct probe of the temperature diffusion, we experimentally retrieve the Hamiltonian of a thermal lattice, and observe the emergence of topological edge decays within the gap of bulk decays. Our results show that such edge states exhibit robust decay rates, which are topologically protected against disorders. This work constitutes a thermal analogue of topological insulators and paves the way to exploring defect-immune heat dissipation.**

The topology of crystal's band structures has gained increasing attentions in condensed matter physics since the 1980s [1-4]. The topological band theory has revolutionized our understanding on classifications of matters and, as a consequence, many fundamentally new topological electronic materials have been discovered, including topological insulators [5,6], Chern insulators [7,8], and so on. In parallel, the analogy between the quantum mechanical and classical waves has inspired the generalization of numerous concepts in condensed matter physics to the classical-wave systems such as electromagnetic, acoustic, and mechanical wave systems. Intuitively, one can transform the governing equations of classical waves (e.g., Maxwell equations for electromagnetic waves) to the Hamiltonian formulation. Following this methodology, topological phases initially proposed for quantum-mechanical waves have been recently realized in various classical-wave systems [9-14], enabling many practical applications such as topological lasers [15-18], robust optical delay lines [19], and high-quality on-chip communication [20].

Another important natural existing physical dynamics, known as diffusion, has been widely studied in the context of heat transfer [21], Brownian motion [22], etc. Unlike fields in wave systems (quantum mechanical or classical), diffusive fields are not governed by "frequency" and thus do not have a phase in time. These fundamental differences between wave and diffusive fields prevent the direct extension of the band theory and topological concepts from wave to diffusion systems. A recent theoretical attempt has shown that the concept of topological phases can be applied to heat diffusion, which is manifested as topologically robust interfacial thermal decay rates within the gap of bulk decay rates [23]. Yet, such topological states of diffusion have not been experimentally observed.

Here, we transfer the notion of topological phases to diffusion dynamics and explore experimentally the topological heat diffusion. Fundamentally different from wave systems that are usually Hermitian, the diffusion system is intrinsically anti-Hermitian, and thus the corresponding eigenvalue $\omega$ is purely imaginary. Such an imaginary eigenvalue corresponds to a decay rate $\gamma$ as $\gamma = -\text{Im}(\omega)$ [24]. In other words, for diffusive dynamics, the topological band theory applies to the decay rates, rather than the frequencies as in wave systems. The topological interfacial diffusion displays unique temporal decays within the decay-rate gap, where the bulk diffusion is forbidden. Moreover, since the phase of eigenfunctions in the topological diffusion system is absent, this remarkable property enables us to readily retrieve the Hamiltonian via direct measurement of the temperature distribution in time. We highlight that the experimental retrieval of the Hamiltonian is fundamentally important in the investigation of topological physics, but has never been directly accomplished so far.

For the practical implementation of our topological diffusion system, we construct a one-dimensional thermal lattice consisting of periodic aluminium disks (Aluminium alloy 6061-T6) with radius $r=1$ cm connected via aluminium channels, with height $h=1$ cm and lattice constant $a=7$ cm (see the design and photograph of the sample in Fig. 1a-b). The straight and meandering channels are judiciously designed in our experiment to control the thermal diffusivity (denoted as $D_1$ or $D_2$) between two disks. The structured channels have geometric parameters $d=2$ mm, $w=1$ mm, $g=1.3$ mm (all the gap widths $g$ in each twisted channel are equivalent).

The demonstrated topological thermal lattice is an excellent thermal analogue of Su-Schrieffer-Heeger (SSH) model that initially describes electrons hopping on an atomic chain (see the upper panel of Fig. 1a). The disk in the thermal lattice corresponds to the atom in the original SSH model while staggered diffusivities $D_1$ and $D_2$ between disks mimic staggered hopping amplitudes $t_1$ and $t_2$ between atoms. Very recently, a theoretical study [23] showed that, in a lattice structure similar to our design, the governing diffusion equation could be discretized as (see derivation details in supplementary section S1),

$$\partial \overline{\psi}(t) / \partial t = \mathrm{i} H \overline{\psi}(t),  \qquad (1)$$

where $\overline{\psi}(t)$ is the discretized temperature field as a function of time $t$. Since each thermal disk only couples to the two nearest neighbouring ones, the corresponding Hamiltonian $H$ of the system is tridiagonal. As a result, the designed thermal topological lattice with $N$ disks depicted in Fig. 1(b) has an $N \times N$ Hamiltonian given by

$$H = (-i)\begin{bmatrix} D_1+D_2 & -D_1 & 0 & 0 & \cdots & -D_2 \\ -D_1 & D_1+D_2 & -D_2 & 0 & \cdots & 0 \\ 0 & -D_2 & D_1+D_2 & -D_1 & \cdots & 0 \\ 0 & 0 & -D_1 & D_1+D_2 & \cdots & 0 \\ \vdots & \vdots & \vdots & \vdots & \ddots & \vdots \\ -D_2 & 0 & 0 & 0 & \cdots & D_1+D_2 \end{bmatrix}$$, where $D_1$ and $D_2$ denote the intra-diffusivity and inter-diffusivity, respectively, normalized to $(a/2)^2$ (see derivation details in supplementary section S2). Note that values of $D_1$ and $D_2$ depend on the geometry of the unit cell e.g., $D_1 > D_2$ corresponds to the unit cell with a straight intra-cell channel and a meandering inter-cell channel (as indicated by the region enclosed by solid curve in Fig. 1(b)), while $D_1 < D_2$ corresponds to the unit cell with meandering intra-cell channel and straight inter-cell channel (as indicated by the region enclosed by dashed curve in Fig. 1(b)). In this way, the diffusion equation is related to the well-known SSH model initially proposed for quantum systems. Since both $D_1$ and $D_2$ are real numbers in our case, $H$ is an anti-Hermitian version of Hamiltonian in the original SSH model, and its eigenvalues $\omega$ are purely imaginary.

We first measure the bulk states in the topological thermal lattice with the periodic boundary condition. As shown in Fig. 1b-c, the fabricated sample has 12 unit cells in total. An igniter is used to locally heat up an individual site numbered as $j_x$ of the designed structure. To obtain a set of temperature fields $\bar{T}_{j_x}$ on the structure, a thermal camera (FLIR T620) is employed to record the transient heat transfer in the sample at different moments. We implement 24 groups of measurement by heating up each site in sequence to construct the temperature matrix as $T(t) = \begin{bmatrix} \bar{T}_1 & \bar{T}_2 & \cdots & \bar{T}_{24} \end{bmatrix}$. The transient heat transfer process follows $\bar{T}(t) = e^{-iHt}\bar{T}(0)$, and, therefore, the real-space Hamiltonian can be retrieved from the measured temperature matrix as

$$H_{\exp} = (-i)\frac{\ln\left[T(t_2)T^{-1}(t_1)\right]}{t_2 - t_1}, \tag{2}$$

where $T(t_1)$ and $T(t_2)$ correspond to the spatial distribution of the temperature at different measurement moments $t_1$ and $t_2$, respectively. Thanks to the absence of phase for diffusion field, the retrieval procedure of the real-space Hamiltonian is practically feasible in our case. In comparison, the measurement of the real-space Hamiltonian is fundamentally challenging in wave systems, whose phase varies rapidly over time. The measured 24 by 24 real-space Hamiltonian $H_{exp}$ is provided in supplementary databases.

By solving the eigenvalues of the retrieved Hamiltonian $H_{exp}$, we obtain the spectrum of the thermal decay rates, and observe a complete decay-rate gap opened up between 1.67 and 1.85 min$^{-1}$(see the black dots in Fig. 2a). Moreover, the retrieved Hamiltonian enables us to determine the topological invariants of the decay-rate bands (see details in supplementary section S4). We have obtained the topological invariants of two types of unit cells in our experiments, i.e., the one with $D_1>D_2$, and the other with $D_1<D_2$. As shown in Fig. 2, although the spectra of decay rates for two unit cells are identical, the trajectories of the endpoints of the pseudospin vector $\overline{P}=P_x\hat{x}+P_y\hat{y}$ are different, i.e. when $D_1>D_2$ ($D_1<D_2$), the trajectory does not (does) wind around the origin, corresponding to the topologically trivial (nontrivial) phase. Here, the pseudospin vector $\overline{P}$ is determined by Bloch wave functions retrieved from the measured Hamiltonian (see calculation procedures in supplementary section S4), and the winding number equals to the loop number that its end point encircles the origin as the wavevector spans the entire Brillouin zone. Since our lattice preserves sublattice symmetry, the winding number 0 (1) corresponds to a Zak phase of 0 ($\pi$). Our experimentally retrieved results are consistent with that using the SSH model, as shown in Fig. 2.

Topological edge states arise at the domain wall between two topologically distinct thermal lattices. To illustrate this point, we experimentally probe the edge states at two types of domain walls, i.e. domain wall *A* and domain wall *B*, as depicted in Fig. 3a. In the domain wall *A* (*B*), the trivial and nontrivial domains are connected via a meandering (straight) channel which mimics weak (strong) coupling (Fig. 3b). From the measured real-space Hamiltonian, we observe two edge states (denoted as edge state I and II in the following discussion) emerging inside the decay-rate gap of bulk modes (see Fig. 3c). Edge state I (II) shows a symmetric (asymmetric) mode profile at domain wall *A* (*B*), as depicted in Fig. 3d. Owing to the emergence of these edge states, the domain-wall temperature decays at a different rate from that in the bulk. To verify this phenomenon experimentally, we excite edge and bulk states based on the mode distribution (Fig. 4f-h), e.g. edge state I (II) is excited by heating up site 12 (site 1) in the vicinity of the corresponding domain wall, while the bulk states are excited by heating up sites 3 and 4 in the bulk. Then, the decay rate of each mode is determined by measuring the temperature of the heated site. Exponential functions are used to fit the temperature evolution to reveal different decaying processes of edge and bulk states (Fig. 3e). Specifically, a single exponential function is sufficient to fit the temperature evolution of edge states (with the corresponding decay rate 0.52 min$^{-1}$ for edge state I and 0.91 min$^{-1}$ for edge state II). In other words, the edge states, once excited, decay at a constant rate in excellent agreement with the theoretical prediction in Fig. 3c. In sharp contrast, two exponential functions are required to successfully fit the temperature evolution of domain sites, with corresponding decay rate 2.10 min$^{-1}$ and 0.29 min$^{-1}$ as the temperature decays rapidly at the beginning and more gently as time evolves. Such instability of the decay rate results from the excitation of multiple bulk modes in the bulk. Note that the excitation of a specific bulk state requires precisely assigning initial temperature to each site according to the corresponding eigenvector, while in our experiment, the implemented source profile shown in Fig. 3h is a combination of multiple eigenvectors (see mode decomposition in Fig. S8). The eigenvectors in the upper band of the decay rates are responsible for the fast decay in the early stage, while the eigenvectors in the lower band dominate in the slow decay after a while.

We also highlight that the two edge states are topologically protected and are thus robust against disorders. As our numerical calculations demonstrate that the effective diffusivity of the thermal channel is highly dependent on its height (see Fig. S9), we introduce a global disorder to the topological thermal lattice by randomly changing the height of each thermal channel. The measured results prove that the decay rate of edge states is robust against the disorders (see domain-wall temperature evolution and corresponding field distribution in Fig. 4). This robustness is also corroborated by the experimentally retrieved spectrum of decay rates: the disorder dramatically changes the decay rates of the bulk states, e.g. the decay-rate gap reduces from 0.18 to 0.10 in the presence of 30% disorders, whereas the topological edge states (i.e., states I and II) are immune to perturbations in diffusivity (see more discussions in supplementary section S8).

In conclusion, our work provides the first experimental demonstration of nontrivial topological phase and topological edge state in a diffusion system. Strikingly, our results hint that the notion of topological phases can apply to a general class of diffusion systems. Meanwhile, the powerful tool of topological analysis shown here could inspire future research on thermal functional materials with different types of topological phases in one, two, and three dimensions, including 2D and 3D topological insulators, Weyl semimetals, and high-order topological insulators [25]. The demonstrated topological states could find applications in temperature management and thermal information processing with topologically enabled robustness.

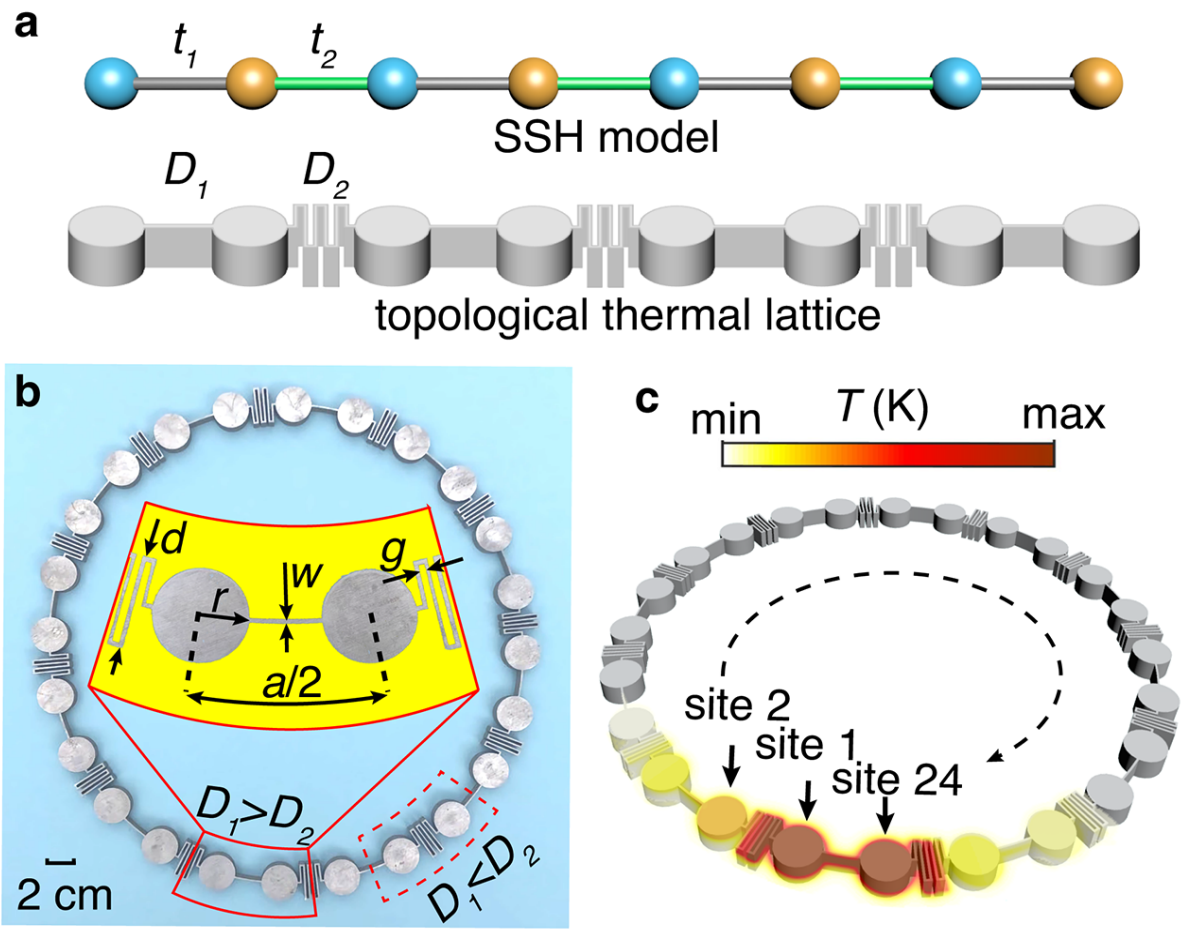

**Figure 1 | Design of a topological thermal lattice with periodic boundary condition**. (**a**) Comparison between the SSH model and the topological thermal lattice. The diffusivity of the thermal field between two disks is determined by the structured channel. The staggered diffusivities $D_1$ and $D_2$ of thermal field in the thermal lattice correspond to the staggered hopping amplitudes $t_1$ and $t_2$ in the SSH model. (**b**) Photograph of the topological thermal lattice consisting of 12 unit cells and 24 sites. Two types of unit cells are highlighted: the one enclosed by the solid curve, and the other enclosed by the dashed curve. (**c**) Schematics of the temperature distribution in the thermal lattice when site 1 of the structure is heated. The sites are numbered clockwise.

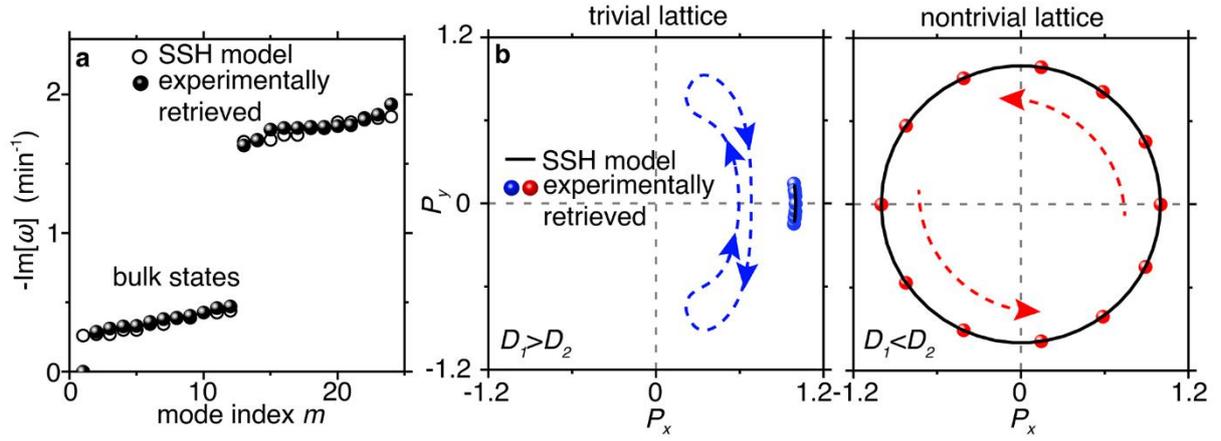

**Figure 2 | Experimentally retrieved eigenvalues of decay rates and topological invariants**. (**a**) Spectrum of decay rates retrieved from measured Hamiltonian. (**b**) The paths of the endpoints of the pseudospin vector $\overline{P}=P_x x + P_y y$ on the $P_xOP_y$ plane (i.e., the equatorial plane of the Bloch sphere) for the lower band of decay rates. The winding number equals the loop number of the endpoints of pseudospin vector that encircle the origin, when $k_x$ goes through the Brillouin zone from 0 to $2\pi$. The evolution trajectory of endpoints of the pseudospin vector is indicated by dashed arrows. In the left panel of (b), the studied unit cell has $D_1>D_2$ while in the right panel of (b), the studied unit cell has $D_1<D_2$.
10

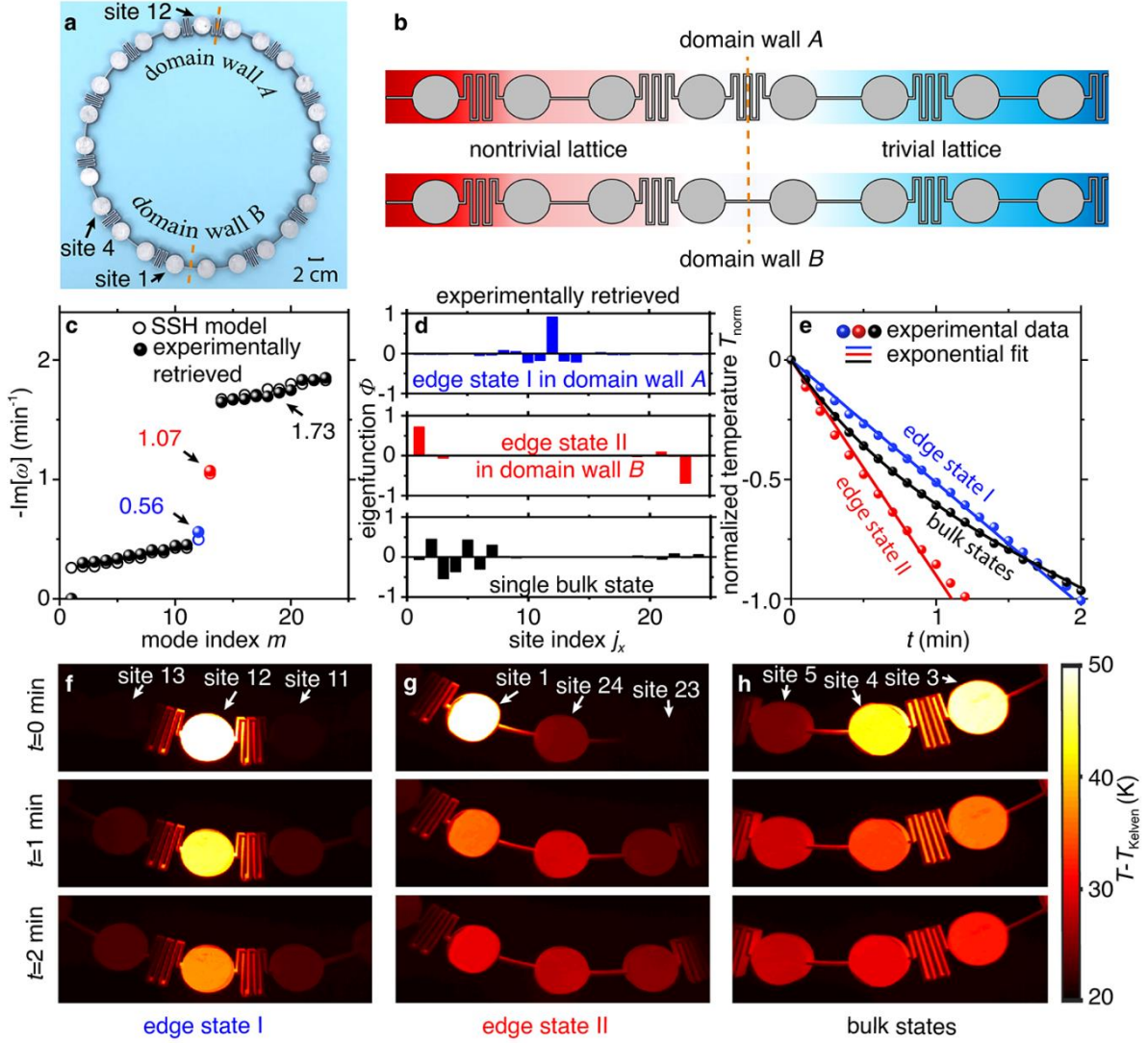

**Figure 3 | Observation of topological edge states.** (**a**) Photograph of the topological thermal lattice with domain walls. The dash lines indicate the positions of domain walls. (**b**) Schematics of domain wall *A* and *B*, respectively. (**c-d**) Retrieved decay-rate bands and mode profiles in the topological thermal lattice. (**e**) Time evolution of normalized temperature for edge states and bulk states. The temperature is normalized with $T_{\text{norm}} = \ln\left[(T - T_f)/(T_i - T_f)\right]$, where $T_i$ is the temperature of heated site at *t*=0 min and $T_f$ is the room temperature. One exponential function with a decay rate of 0.52 (0.91) is sufficient to fit the measured time evolution of temperature for edge state I (edge state II), while two exponential functions with decay rates of 2.10 and 0.29 are required to well fit the time evolution of temperature for bulk states. (**f-h**) Temperature field distribution for edge states and bulk states. The temperature field distributions are recorded at *t*=0 min, *t*=1 min, and *t*=2 min, respectively.

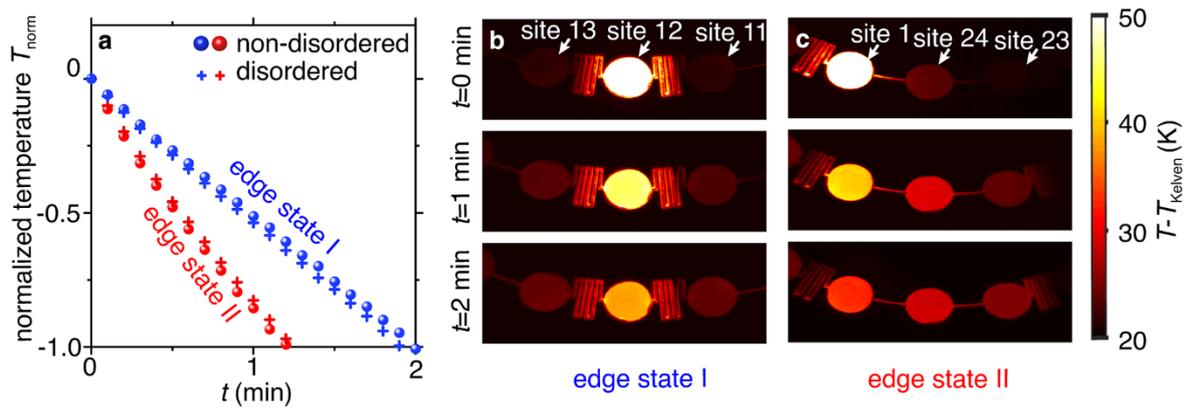

**Figure 4 | Influence of disorders on the topological edge states.** (**a**) Time evolution of normalized temperature for edge states in non-disordered and disordered systems, respectively. (**b-c**) Temperature field distribution for edge states. The temperature field distributions are recorded at *t*=0 min, *t*=1 min, and *t*=2 min, respectively.